\documentclass[12pt,preprint]{aastex}

\usepackage[usenames]{color}
\newcommand{\kms}{\mbox{km s$^{-1}$}}

\slugcomment{Draft June 21 2012}

\shorttitle{Disentangling the circumnuclear environs of Centaurus A: Gaseous Spiral Arms}
\shortauthors{D. Espada et al.}

\begin{document}

\title{Disentangling the circumnuclear environs of Centaurus A: \\
 Gaseous Spiral Arms in a Giant Elliptical Galaxy}

\author{
D. Espada \altaffilmark{1,2}, S. Matsushita \altaffilmark{3,4}, 
A.~B. Peck \altaffilmark{4,5},  
C. Henkel \altaffilmark{6,7}, 
F. Israel \altaffilmark{8},
D. Iono \altaffilmark{9}
}

\altaffiltext{1}{National Astronomical Observatory of Japan (NAOJ), 2-21-1 Osawa, Mitaka, Tokyo 181-8588, Japan; daniel.espada@nao.ac.jp}
\altaffiltext{2}{Harvard-Smithsonian Center for Astrophysics, 60 Garden St., Cambridge, MA 02138, USA}
\altaffiltext{3}{Academia Sinica, Institute of Astronomy and Astrophysics, P.O. Box 23-141, Taipei 10617, Taiwan}
\altaffiltext{4}{Joint ALMA Observatory, Alonso de Cordova 3107, Vitacura - Santiago, Chile}
\altaffiltext{5}{NRAO, 520 Edgemont Rd, Charlottesville VA 22903, USA}
\altaffiltext{6}{Max-Planck-Institut f{\"u}r Radioastronomie, Auf dem H{\"u}gel 69, 53121 Bonn, Germany}
\altaffiltext{7}{Astron. Dept., King Abdulaziz University, P.O. Box 80203, Jeddah, Saudi Arabia}
\altaffiltext{8}{Sterrewach Leiden, Leiden University, Niels Bohrweg 2, 2333 CA Leiden, The Netherlands}
\altaffiltext{9}{Nobeyama Radio Observatory, NAOJ, Minamimaki, Minamisaku, Nagano, 384-1305, Japan}

\begin{abstract}
We report the existence of spiral arms in the recently formed gaseous and dusty disk of the closest giant elliptical, NGC~5128 (Centaurus A), using high resolution $^{12}$CO(2--1) observations of the central 3\arcmin\ (3~kpc) obtained with the Submillimeter Array (SMA). This provides evidence that spiral-like features can develop within ellipticals if enough cold gas exists.
 We elucidate the distribution and kinematics of the molecular gas in this
 region with a resolution of 4\farcs4 $\times$ 1\farcs9 (80~pc $\times$ 40~pc). 
 The spiral arms extend from the circumnuclear gas at a radius of 200~pc to at least 1~kiloparsec. 
 The general properties of the arms are similar to those in spiral galaxies: they are trailing, their width is $\sim$ 500 $\pm$ 200~pc, and the pitch angle is $\sim$20$\rm ^o$.
From independent estimates of the time when the HI-rich galaxy merger occurred, we infer that the formation of spiral arms occured on a time scale of less than $\sim$10$^8$~yr. The formation of spiral arms increases the gas density and thus the star formation efficiency in the early stages of the formation of a disk.

\end{abstract}


\section{Introduction}
\label{sec1}

Some ellipticals show rotating disk-like features of gas and dust, expected to form as a result of gas accretion from the intergalactic medium, cannibalization of other galaxies, or re-accretion of material originally expelled during the major merger event that resulted in the formation of the elliptical itself \citep[e.g.]{1997SSRv...81....1H,2002AJ....124..788Y,2011MNRAS.417..882D,2011MNRAS.414..940Y}.  %The small percentage ($\sim$ 22\%) of early type galaxies  with significant gas and dust can be classified into those with settled disks (from the shape of spectral profiles) and those with unsettled structures \citep{2011MNRAS.414..940Y}.
%The ellipticals are periodically replenished with gas and dust by \citep[e.g.]{1997SSRv...81....1H}. 
\citet{1993ApJ...419..544S} proposed an evolutionary sequence of disk formation in dust lane ellipticals via accretion, from irregularly distributed gas to more evolved and settled disks. However, our understanding of the properties and evolution of such relatively settled disks is poorly understood yet, 
partly because most observations of the gas in elliptical galaxies are hampered by poor angular resolution and/or sensitivity. In particular, it is not known if these newly created disks in elliptical galaxies are uniform once settled, or whether deviations from axisymmetry such as prominent spiral structures can develop, affecting the subsequent star formation patterns, and the overall evolution of the disk itself.

\begin{figure}
\centering
\includegraphics[width=18cm]{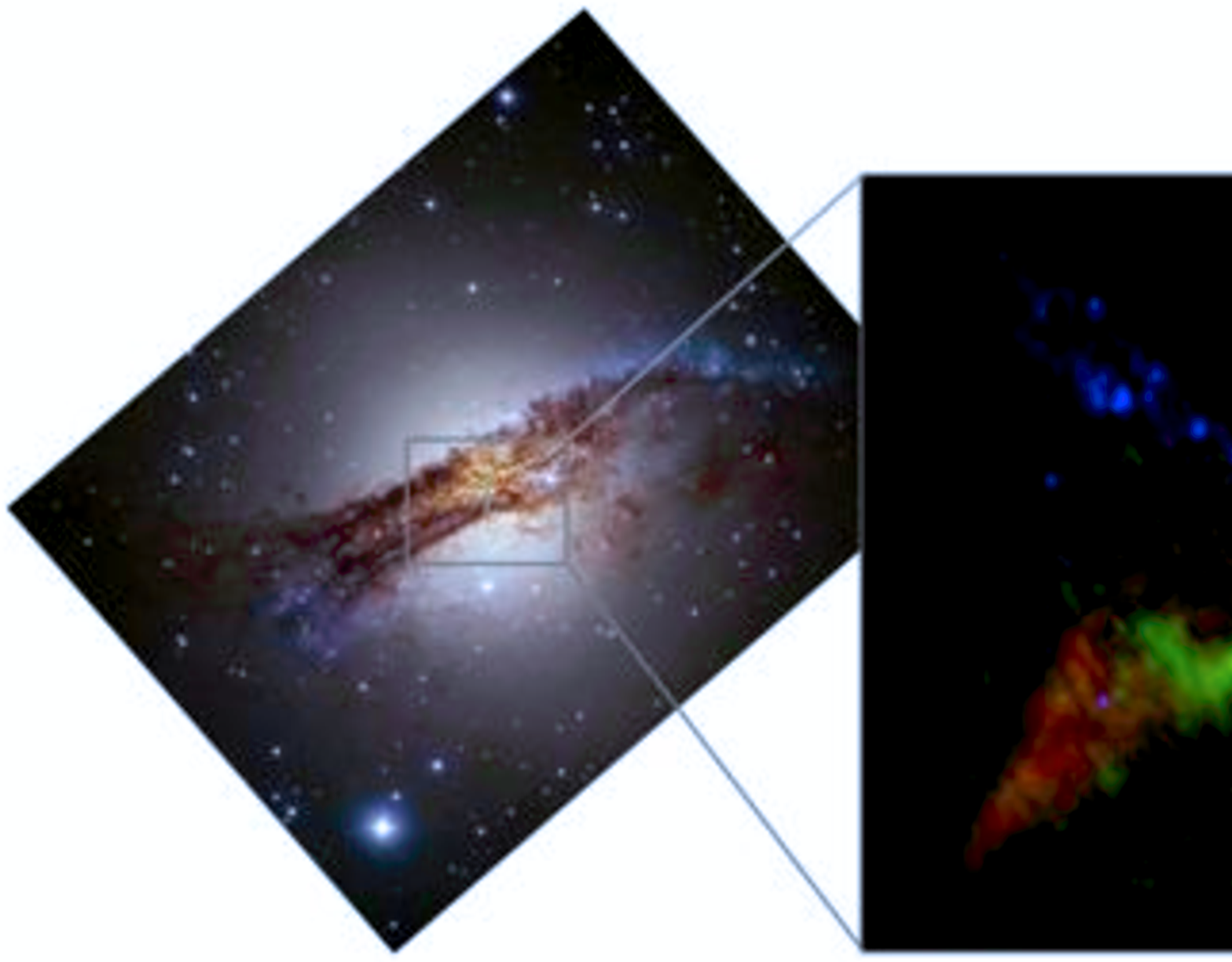}
\caption[Comparison of molecular gas (SMA CO(2--1) map, Fig. 1a), PAH and dust emission (mid-IR and X-ray jet]{
\emph{Left)} Optical image of Centaurus A, showing its prominent dust lane (ESO/IDA/Danish 1.5 m/R. Gendler, J.-E. Ovaldsen \& S. Guisard, ESO). \emph{Right)} The molecular gas as traced by our SMA CO(2--1) observation (green), PAH and dust emission at 8 $\mu$m observed by Spitzer  \citep[red,][]{2006ApJ...645.1092Q},  and the Chandra X-ray observations of the jet (blue, NASA/CXC/M. Karovska et al.). Note that the spiral and the 8$\mu$m emission in this panel are within the optical dust lane visible in the left panel. The most prominent features of the $^{12}$CO(2--1) emission are consistent with those of the $8~\mu$m emission, but the CO shows the spiral arms. There is a CO absorption feature toward the unresolved compact continuum component located toward the AGN \citep{2010ApJ...720..666E}. Note that the jet is nearly perpendicular to the circumnuclear molecular gas at the base of the spiral arms, but not to the outer component.  
\label{fig1}}
\end{figure}

At a distance of D $\simeq$ 3.8~Mpc \citep{2010PASA...27..457H} (where 1\arcsec\ roughly corresponds to 18~pc), NGC~5128 (hereafter Cen~A) is the closest giant elliptical \citep{1976ApJ...208..673V,1998AARv...8..237I,2012AJ....143...84H}
with a prominent dust lane along its minor axis \citep{1998AARv...8..237I,2012AJ....143...84H} (see Figure~\ref{fig1}, left). 
This dust lane contains large amounts of gas and dust, as traced by H$\alpha$ \citep{1992ApJ...387..503N}, mid-IR \citep{2006ApJ...645.1092Q}, \ion{H}{1} \citep{1990AJ.....99.1781V,2010AA...515A..67S}, and CO \citep{1992ApJ...391..121Q,1993AA...270L..13R,2009ApJ...695..116E}. 

The \ion{H}{1} emission still shows unsettled gas at r$>$6~kpc, including tail/arm like structures, and it is estimated that the merger event occurred within a timescale of only 0.3~Gyr  \citep{2010AA...515A..67S}. The gas in molecular phase and the dusty component inside the inner few kiloparsecs are well settled within a warped disk of about 7\arcmin\ with an inclination of $i$ $\sim$ 70$^{\rm o}$ \citep{2006ApJ...645.1092Q}, although its detailed distribution at scales of a few hundreds of parsec has hitherto been poorly understood. 

The most popular model to reproduce the large scale disk structure is that of an homogenous warped and thin disk (e.g. \citealt{1992ApJ...387..503N,1992ApJ...391..121Q,2006ApJ...645.1092Q,2009AA...502L...5K,2010AA...515A..67S,2010PASA...27..396Q}). According to these models, the disk crosses the line of sight at two galactocentric radii. One of the most remarkable features in favor of this model is that in projection it can reproduce the parallelogram structure seen in the IR, having a major side of $\sim$3\arcmin\ (3~kpc) along a PA = 120$\rm ^o$ \citep{2006ApJ...645.1092Q}. 

However,  the inner 1 kiloparsec of Cen~A (up to a few 100 pc from the nucleus) is very complex. 
\citet[][hereafter Paper I]{2009ApJ...695..116E} report $^{12}$CO(2--1) SMA observations of its nuclear region in a single pointing, with a resolution of 6\farcs0 $\times$ 2\farcs4 (100 $\times$ 40 pc), revealing the detailed distribution and kinematics of the molecular gas in the central inner kpc, including a compact circumnuclear disk of molecular gas surrounding the AGN, as well as outer molecular gas partly associated with the parallelogram structure.
While the warped disk models reproduce the observed features (distribution, kinematics, dust lane appearance)
at a large scale, Paper I shows that the molecular gas distribution in the inner kpc is not sufficiently in agreement with these models. A weak bi-symmetric potential was proposed in Paper I to explain the deviations with respect to this model. 

We present new observations using the Submillimeter Array (SMA\footnote{The Submillimeter Array is a joint project between the Smithsonian Astrophysical Observatory and the Academia Sinica Institute of Astronomy and Astrophysics, and is funded by the Smithsonian Institution and the Academia Sinica.}; \citealt{2004ApJ...616L...1H}), that allow us to shed more light into the complex few inner kiloparsecs. The resolution of the mosaic we present here is a factor of 45 higher than existing $^{12}$CO(2--1) maps covering the dust lane, with a resolution of $23\arcsec$ (or $\sim380$~pc) \citep{1993AA...270L..13R}. Our mosaic covers an area three times larger than previously published interferometric CO maps (Paper I) to fully cover the entire parallelogram structure.

\section{SMA $^{12}$CO(2--1) Observations and Data Reduction}
\label{sec2}

Cen~A was observed at 1.3~mm using the SMA with 7 antennas on April 22 and 26 2008. 
The field of view is characterized by a Half Power Beam Width (HPBW) of the primary beam of an SMA antenna of $52\arcsec$ (0.9~kpc). We observe a 5 pointing mosaic to cover 156 $\times$ 52\arcsec (or 2.6 $\times$  0.9~kpc).
The digital correlator was configured with 6144 channels (2~GHz bandwidth), resulting in a velocity resolution of about 0.5~\kms.  The receivers were tuned to the redshifted $^{12}$CO(2--1) ($\nu_{\rm rest}$ = 230.538~GHz) emission line in the upper sideband (USB), using $V_{\rm LSR}$ = 550~\kms . Note that velocities are expressed throughout this paper with respect to the LSR using the radio convention. 
We used R.A.\ = $13^{\rm h}25^{\rm m}27\fs6$ and Dec.\ = $-43\arcdeg 01 \arcmin 08\farcs8$ (J2000) as our phase center (AGN position R.A.\ = $13^{\rm h}25^{\rm m}27\fs615$ and Dec.\ = $-43\arcdeg 01 \arcmin 08\farcs805$, \citealt[][]{1998AJ....116..516M}).  
The maximum elevation of the source at the SMA site is $\simeq 27\arcdeg$, forcing us to observe only under very stable atmospheric conditions, with zenith opacities of typically $\tau_{225}\sim0.10$.  In order to have a beam shape as close to circular as possible we used a compact configuration with longer N--S baselines. Minimum and maximum projected baselines were 6~m and 108~m. The maximum angular scale observable with the shortest baselines is 25\arcsec, which considerably limits the amount of missing flux to emission with angular scales larger than this. 

The two tracks were reduced independently.
The editing and calibration of the data were done with the SMA-adapted MIR software\footnote{MIR is a software package to reduce SMA data based on the package originally developed by Nick Scoville at Caltech.  See \url{http://cfa-www.harvard.edu/~cqi/mircook.html}}. 3C273 and 3C279 were used for passband calibration.  An initial gain calibration was performed using J1337-129, which is at an angular distance of $30\arcdeg$ from the target.  We confirm that the continuum emission toward Cen~A was found to be unresolved (Paper I).  The gain calibration for the five pointings was then refined using the averaged line-free channels in Cen~A itself (central pointing). 
Callisto and Ganymede were used as absolute flux calibrators. Overall, we estimate the absolute flux uncertainties on the order of 10\%, from the comparison of both datasets. Finally we combined both tracks.

 An image of the $^{12}$CO(2--1) emission was produced using MIRIAD \citep{1995ASPC...77..433S}.  A careful subtraction of the continuum was done using line-free channels with the task \verb!UVLIN!.  The data were cleaned using \verb!MOSMEM! with uniform weighting. 
The synthesized beam is $4\farcs4 \times 1\farcs9$ ($80 \times 40$~pc) with a major axis P.A.\ = $25.3\arcdeg$.  
The task \verb!MOMENT! was used to calculate the $^{12}$CO(2--1) integrated flux density distribution (2$\sigma$ clipping) and the intensity-weighted velocity field distribution (clipped at 3$\sigma$). The rms noise level of the integrated intensity map is 6~Jy~beam$^{-1}$~km~s$^{-1}$.  
Figure~\ref{fig3} shows the SMA  $^{12}$CO(2--1) integrated intensity map and velocity field along the inner 3\arcmin~($\sim$3~kpc).

\section{Spiral arms in a Giant Elliptical Galaxy}
\label{sec3}

\begin{figure}
\begin{center}
\includegraphics[width=12cm]{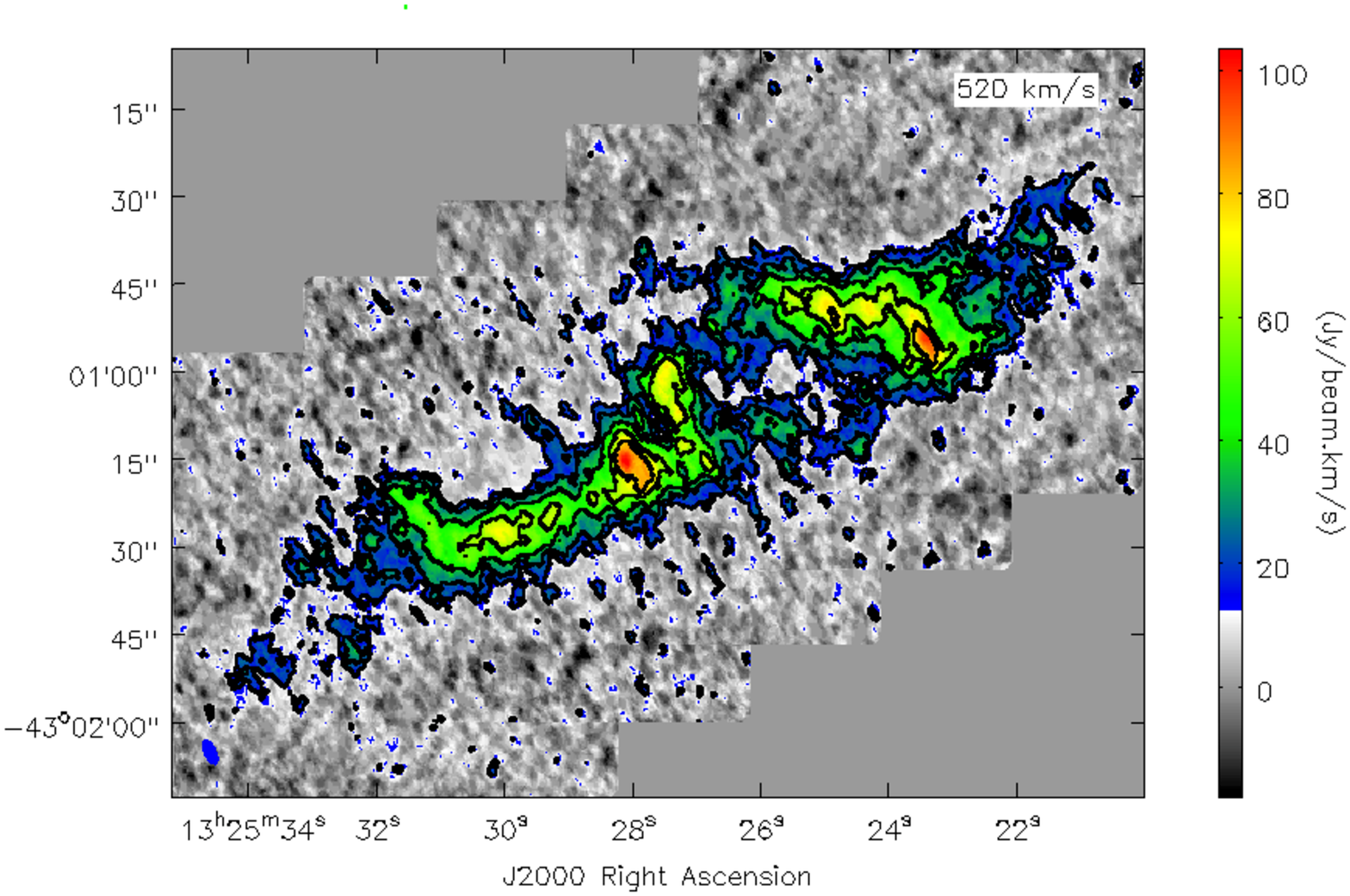}
\includegraphics[width=12cm]{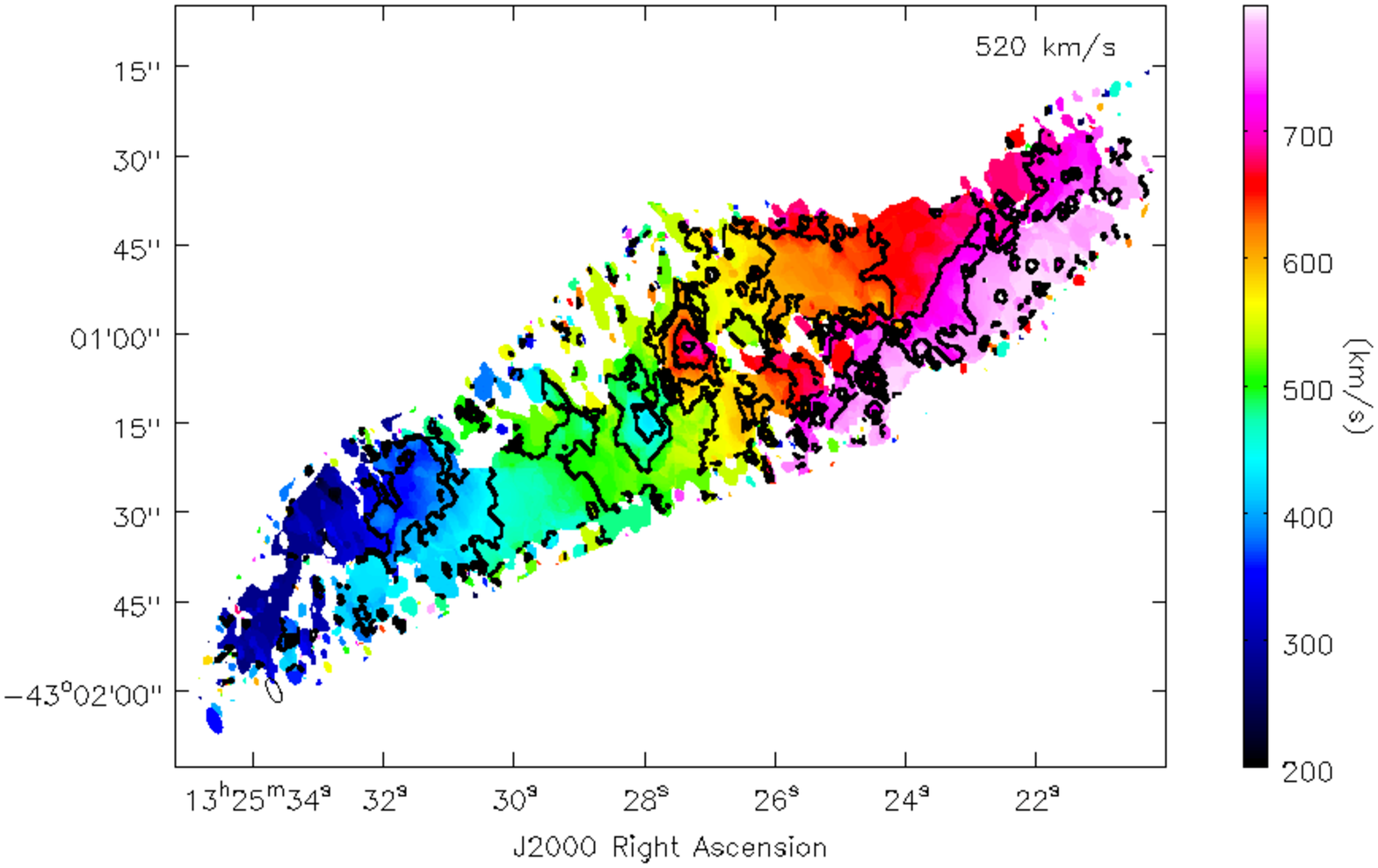}
\end{center}
\caption{\emph{Top)} $^{12}$CO(2--1) integrated intensity map. Color scale ranges from -20 to 103~Jy~beam$^{-1}$~km s$^{-1}$.  The rms of the CO image is 6~Jy~beam$^{-1}$~km s$^{-1}$. \emph{Bottom)} $^{12}$CO(2--1) (intensity weighted) velocity field map.  Contours are placed every 50~km~s$^{-1}$, from $200 - 800$~km~s$^{-1}$.  The color scale ranges from 400 \kms\ up to 700 \kms . The size of the synthesized beam is shown in the lower left corner (dark blue filled ellipse). 
\label{fig3}}
\end{figure}

From the $^{12}$CO(2--1) emission distribution in Figure~\ref{fig3}, we confirm that the molecular gas is preferentially located along two filamentary structures resembling spiral arm like features to the SE and NW of the circumnuclear gas (galactocentric radius $<$ 200~pc and position angle P.A. = 155$\arcdeg$, Paper I).
The larger field of view of our observations demonstrates that these filamentary structures extend at least 1\farcm0, and they are curved toward the NE and SW, respectively, as projected distance from the nucleus increases. 
The approaching and receding sides of the disk are on the E and W, respectively, and the velocity range spans about 600~\kms, from 200 to 800~\kms.

 The near side of the disk at $r$ $<$ 1.3~kpc is in the South \citep[][]{2010PASA...27..396Q}. From the morphology of the bisymmetric spiral arm like features and the velocity field in Figure~\ref{fig3}, we infer that they are trailing (convex side advances).
The projected angular widths of the spiral arm like features are $\sim$10\arcsec . Adopting an inclination of 70$^{\rm o}$  \citep{2010PASA...27..396Q}, their widths are 500~$\pm$~200~pc. %Interestingly, both the trailing spiral arm like features and the widths are in agreement with the properties of arms in disk galaxies. 

\begin{figure}
\centering
\includegraphics[width=10cm]{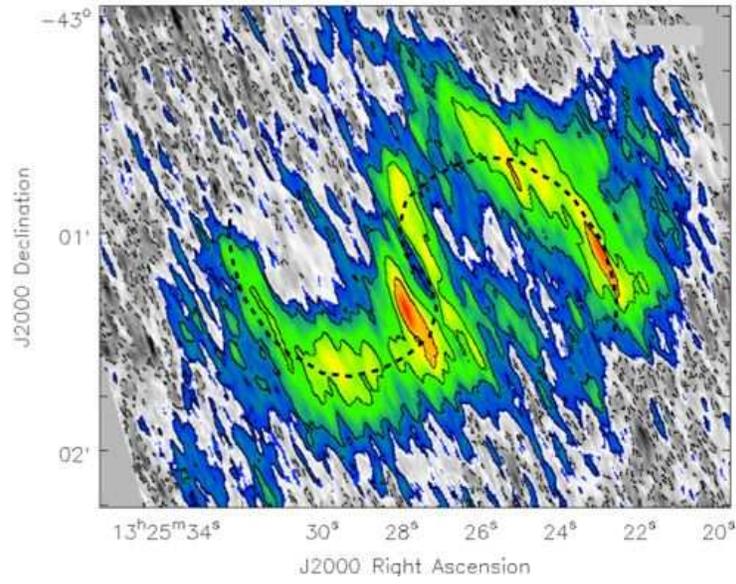}
\caption[Deprojected CO(2--1) distribution]{The deprojected CO(2--1) distribution and a logarithmic spiral pattern with pitch angle $\phi$ = 20$^{\rm o}$ (black dashed line). \label{fig2}}
\end{figure}

The shape of galactic spiral arms is usually logarithmic in nature and independent of scale \citep[e.g.][]{1998MNRAS.299..685S}. 
One of the best geometric measures to represent spiral arms is the pitch angle, $\phi$, defined as the angle between the line tangent to a circle and the line tangent to a logarithmic
spiral at a specified radius. We present in Figure~\ref{fig2} the deprojected CO(2--1) distribution, as well as a best fit representation of a logarithmic spiral pattern with pitch angle $\phi$ = 20$^{\rm o}$.  We used a set of models of bisymmetric spiral arms with an inclination of 70$^{\rm o}$. We varied the pitch angles in steps of 10$^{\rm o}$ from 10 to 60$^{\rm o}$, and found that $\phi$ = 20$^{\rm o}$ fitted best.  This pitch angle is  within the expected range for late type spirals \citep{1981AJ.....86.1847K}  for a maximum rotation velocity for Cen~A of $V$ $\simeq$ 300~\kms . 

\begin{figure}
\centering
\includegraphics[width=12cm]{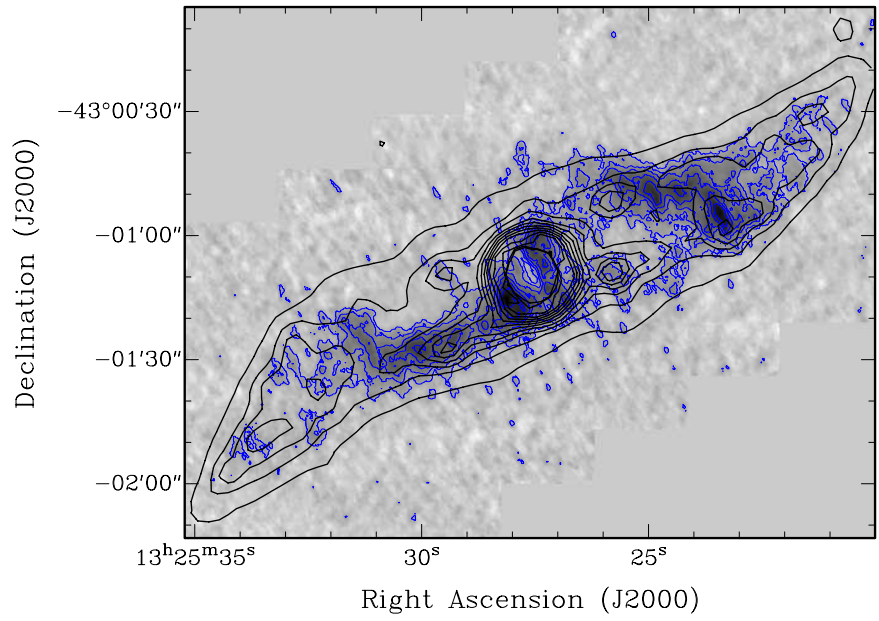}
\caption[Deprojected CO(2--1) distribution]{Comparison of Spitzer 24$\mu$m emission (contours) and CO(2--1) emission (grey scale). \label{fig4}}
\end{figure}

In Figure~\ref{fig4} we show a comparison between the CO(2--1) and mid-IR 24$\mu$m emission obtained with Spitzer \citep{2006ApJ...645.1092Q}, with resolutions of about 6\arcsec, which complements Figure~\ref{fig1} with its mid-IR 8$\mu$m emission. In general the molecular gas distribution is similar to that of the dust emission. 

As mentioned previously, a warped and thin disk model is usually used to reproduce the observed properties at large scales ($>$1~kpc), especially the 3\arcmin\  parallelogram structure seen in dust emission \citep{2006ApJ...645.1092Q,2010PASA...27..396Q}. However, the morphology of CO(2--1) emission resembles better spiral arms rather than a parallelogram structure. This rules out the single contribution of a homogeneous disk, otherwise we would also see filaments to the NE and SW of the circumnuclear gas, as the warped and thin disk model predicts that they are of comparable brightness as the observed NW and SE filaments (see for example Figure~4 in \citealt{2006ApJ...645.1092Q}). 
This cannot be explained by flux loss due to missing spacings. From the model, these filaments should have sizes less than 10\arcsec, similar compactness than the filaments to the NE an SW, and thus they would have been detected.

Therefore, we interpret the two CO(2--1) filamentary structures as 
spiral arms. Their main properties are similar to those found in disk galaxies: they are trailing, the linear widths are a few hundred parsecs, and the pitch angle is typical of spiral galaxies. 
This provides evidence that spiral arms can develop within a giant elliptical if enough cold gas exists.  However, note that the molecular gas properties in the spiral arms might be considerably different to those in spiral galaxies, as it coexists within a triaxial potential and the stellar surface density changes more abruptly with radius in the case of a giant elliptical.

\section{Discussion}

A consequence of the compression produced by these spiral arms is that it is expected to trigger star formation (SF) on the leading edge. This is consistent with abundant SF traced by Pa$\alpha$ \citep{2000ApJ...528..276M} for example, coincident at least with the southern molecular spiral arm (nearest part and thus likely the least obscured). 
The spiral features seen in the $^{12}$CO emission is also associated with the distribution of certain tracers observed in poorer angular/spectral resolution maps, such as the SCUBA 450$\mu$m emission map in \citet{2002ApJ...565..131L}, as well as the pure rotational line of molecular hydrogen H$_2$ (J=2--0) S(0) ($28.22~\mu$m) emission observed with Spitzer/IRS by \citet{2008MNRAS.384.1469Q}.
The molecular hydrogen transition H$_2$ (J=2--0) S(0)  indicates the presence of gas with T $\sim$ 200~K \citep{2008MNRAS.384.1469Q}, which is likely tracing photodissociation regions associated with abundant star formation.

A small perturbation could have triggered this spirality, such as a non-axisymmetric weak potential (Paper I) or a minor merger after the disk was relatively well settled \citep[e.g.][]{2011Natur.477..301P}.
From the \ion{H}{1} structure and kinematics we infer that the formation of spiral arms took place in less than 0.3~Gyr \citep{2010AA...515A..67S}, which is likely the most accurate measure of time at our disposal since \ion{H}{1} is one of the most sensitive components of interaction. 
Although evidence suggests that spiral arms are transient \citep{2011MNRAS.410.1637S,2011ApJ...735..101F}, recent simulations claim that spiral structures may be long lived \citep{2012arXiv1204.0513D}.
3-D self-gravitating models for late type spirals suggest that the pitch angle we find here would correspond to a long-lasting feature, because the density response tends to avoid larger pitch angles \citep{2012ApJ...745L..14P}. 
Similar simulations in the deep potentials of giant ellipticals would be needed to investigate the response of the gas. 

The study of the properties of recently accreted molecular gas in the deep potential wells of elliptical galaxies is a powerful tool in disentangling how galaxies form and evolve.  
Recent smoothed-particle hydrodynamic (SPH) cosmological simulations in the $\Lambda$ Cold Dark Matter (CDM) scenario \citep{2011ApJS..192...18K} are able to reproduce most of the properties of disks \citep{2011MNRAS.410.1391A,2011ApJ...742...76G,2012MNRAS.tmp.2503D} at the current epoch,  including their fine structure morphology, the observed angular momentum, the Tully-Fisher relation, and the SFR to gas surface density ratio (Kennicutt-Schmidt). 
However, to understand galaxy evolution it is essential to anchor these simulations with the actual properties of nascent disk galaxies. The properties of disks, in particular the existence of spiral arms, in these systems which are in the early stages following significant accretion can be compared with numerical simulations. 
Not only can we compare the observed gas distribution and kinematics of recently formed disks to models, but also the effect of SF and Active Galactic Nuclei (AGN) feedback in the disk evolution. 
Moreover, with the advent of ALMA, it is now possible to validate theories of disk formation as they enter a deep gravitational potential, performing similar studies in objects along the evolutionary sequence of accreted gas in dust lane ellipticals \citep{1993ApJ...419..544S},  from systems whose gas is still irregularly distributed to more evolved and settled disks such as in Cen~A.

\acknowledgments{
We thank the referee for useful comments that improved the focus of this Letter. 
This research has made use of NASA's Astrophysics Data System Bibliographic Services, and has also made use of the NASA/IPAC Extragalactic Database (NED) which is operated by the Jet Propulsion Laboratory, California Institute of Technology, under contract with the National Aeronautics and Space Administration. This research was partially supported by a Marie Curie International Fellowship within the 6$\rm ^{th}$ European Community Framework Programme.

}

{\it Facilities:} \facility{SMA}

\begin{thebibliography}{37}
\expandafter\ifx\csname natexlab\endcsname\relax\def\natexlab#1{#1}\fi


\bibitem[{{Agertz} {et~al.}(2011){Agertz}, {Teyssier}, \&
  {Moore}}]{2011MNRAS.410.1391A}
{Agertz}, O., {Teyssier}, R., \& {Moore}, B. 2011, \mnras, 410, 1391

\bibitem[Davis et al.(2011)]{2011MNRAS.417..882D} Davis, T.~A., Alatalo, 
K., Sarzi, M., et al.\ 2011, \mnras, 417, 882 

\bibitem[{{Dom{\'e}nech-Moral} {et~al.}(2012){Dom{\'e}nech-Moral},
  {Mart{\'{\i}}nez-Serrano}, {Dom{\'{\i}}nguez-Tenreiro}, \&
  {Serna}}]{2012MNRAS.tmp.2503D}
{Dom{\'e}nech-Moral}, M., {Mart{\'{\i}}nez-Serrano}, F.~J.,
  {Dom{\'{\i}}nguez-Tenreiro}, R., \& {Serna}, A. 2012, \mnras, 2503

\bibitem[{{D'Onghia} {et~al.}(2012){D'Onghia}, {Vogelsberger}, \&
  {Hernquist}}]{2012arXiv1204.0513D}
{D'Onghia}, E., {Vogelsberger}, M., \& {Hernquist}, L. 2012, ArXiv e-prints

\bibitem[{{Espada} {et~al.}(2009){Espada}, {Matsushita}, {Peck}, {Henkel},
  {Iono}, {Israel}, {Muller}, {Petitpas}, {Pihlstr{\"o}m}, {Taylor}, \&
  {Dinh-V-Trung}}]{2009ApJ...695..116E}
{Espada}, D., {et~al.} 2009, \apj, 695, 116

\bibitem[{{Espada} {et~al.}(2010){Espada}, {Peck}, {Matsushita}, {Sakamoto},
  {Henkel}, {Iono}, {Israel}, {Muller}, {Petitpas}, {Pihlstr{\"o}m}, {Taylor},
  \& {Trung}}]{2010ApJ...720..666E}
---. 2010, \apj, 720, 666

\bibitem[{{Foyle} {et~al.}(2011){Foyle}, {Rix}, {Dobbs}, {Leroy}, \&
  {Walter}}]{2011ApJ...735..101F}
{Foyle}, K., {Rix}, H.-W., {Dobbs}, C.~L., {Leroy}, A.~K., \& {Walter}, F.
  2011, \apj, 735, 101

\bibitem[{{Guedes} {et~al.}(2011){Guedes}, {Callegari}, {Madau}, \&
  {Mayer}}]{2011ApJ...742...76G}
{Guedes}, J., {Callegari}, S., {Madau}, P., \& {Mayer}, L. 2011, \apj, 742, 76

\bibitem[{{Harris} {et~al.}(2012){Harris}, {G{\'o}mez}, {Harris}, {Johnston},
  {Kazemzadeh}, {Kerzendorf}, {Geisler}, \& {Woodley}}]{2012AJ....143...84H}
{Harris}, G.~L.~H., {G{\'o}mez}, M., {Harris}, W.~E., {Johnston}, K.,
  {Kazemzadeh}, F., {Kerzendorf}, W., {Geisler}, D., \& {Woodley}, K.~A. 2012,
  \aj, 143, 84

\bibitem[{{Harris} {et~al.}(2010){Harris}, {Rejkuba}, \&
  {Harris}}]{2010PASA...27..457H}
{Harris}, G.~L.~H., {Rejkuba}, M., \& {Harris}, W.~E. 2010, \pasa, 27, 457

\bibitem[Henkel 
\& Wiklind(1997)]{1997SSRv...81....1H} Henkel, C., \& Wiklind, T.\ 1997, \ssr, 81, 1 

\bibitem[{{Ho} {et~al.}(2004){Ho}, {Moran}, \& {Lo}}]{2004ApJ...616L...1H}
{Ho}, P.~T.~P., {Moran}, J.~M., \& {Lo}, K.~Y. 2004, \apjl, 616, L1

\bibitem[{{Israel}(1998)}]{1998AARv...8..237I}
{Israel}, F.~P. 1998, \aapr, 8, 237

\bibitem[{{Kainulainen} {et~al.}(2009){Kainulainen}, {Alves}, {Beletsky},
  {Ascenso}, {Kainulainen}, {Amorim}, {Lima}, {Marques}, {Moitinho},
  {Pinh{\~a}o}, {Rebord{\~a}o}, \& {Santos}}]{2009AA...502L...5K}
{Kainulainen}, J.~T., {et~al.} 2009, \aap, 502, L5

\bibitem[{{Kennicutt}(1981)}]{1981AJ.....86.1847K}
{Kennicutt}, Jr., R.~C. 1981, \aj, 86, 1847

\bibitem[{{Komatsu} {et~al.}(2011){Komatsu}, {Smith}, {Dunkley}, {Bennett},
  {Gold}, {Hinshaw}, {Jarosik}, {Larson}, {Nolta}, {Page}, {Spergel},
  {Halpern}, {Hill}, {Kogut}, {Limon}, {Meyer}, {Odegard}, {Tucker}, {Weiland},
  {Wollack}, \& {Wright}}]{2011ApJS..192...18K}
{Komatsu}, E., {et~al.} 2011, \apjs, 192, 18

\bibitem[{{Leeuw} {et~al.}(2002){Leeuw}, {Hawarden}, {Matthews}, {Robson}, \&
  {Eckart}}]{2002ApJ...565..131L}
{Leeuw}, L.~L., {Hawarden}, T.~G., {Matthews}, H.~E., {Robson}, E.~I., \&
  {Eckart}, A. 2002, \apj, 565, 131

\bibitem[{{Ma} {et~al.}(1998){Ma}, {Arias}, {Eubanks}, {Fey}, {Gontier},
  {Jacobs}, {Sovers}, {Archinal}, \& {Charlot}}]{1998AJ....116..516M}
{Ma}, C., {et~al.} 1998, \aj, 116, 516

\bibitem[{{Marconi} {et~al.}(2000){Marconi}, {Schreier}, {Koekemoer},
  {Capetti}, {Axon}, {Macchetto}, \& {Caon}}]{2000ApJ...528..276M}
{Marconi}, A., {Schreier}, E.~J., {Koekemoer}, A., {Capetti}, A., {Axon}, D.,
  {Macchetto}, D., \& {Caon}, N. 2000, \apj, 528, 276

\bibitem[{{Nicholson} {et~al.}(1992){Nicholson}, {Bland-Hawthorn}, \&
  {Taylor}}]{1992ApJ...387..503N}
{Nicholson}, R.~A., {Bland-Hawthorn}, J., \& {Taylor}, K. 1992, \apj, 387, 503

\bibitem[{{P{\'e}rez-Villegas} {et~al.}(2012){P{\'e}rez-Villegas}, {Pichardo},
  {Moreno}, {Peimbert}, \& {Vel{\'a}zquez}}]{2012ApJ...745L..14P}
{P{\'e}rez-Villegas}, A., {Pichardo}, B., {Moreno}, E., {Peimbert}, A., \&
  {Vel{\'a}zquez}, H.~M. 2012, \apjl, 745, L14

\bibitem[{{Purcell} {et~al.}(2011){Purcell}, {Bullock}, {Tollerud}, {Rocha}, \&
  {Chakrabarti}}]{2011Natur.477..301P}
{Purcell}, C.~W., {Bullock}, J.~S., {Tollerud}, E.~J., {Rocha}, M., \&
  {Chakrabarti}, S. 2011, \nat, 477, 301

\bibitem[{{Quillen} {et~al.}(2006){Quillen}, {Brookes}, {Keene}, {Stern},
  {Lawrence}, \& {Werner}}]{2006ApJ...645.1092Q}
{Quillen}, A.~C., {Brookes}, M.~H., {Keene}, J., {Stern}, D., {Lawrence},
  C.~R., \& {Werner}, M.~W. 2006, \apj, 645, 1092

\bibitem[{{Quillen} {et~al.}(1992){Quillen}, {de Zeeuw}, {Phinney}, \&
  {Phillips}}]{1992ApJ...391..121Q}
{Quillen}, A.~C., {de Zeeuw}, P.~T., {Phinney}, E.~S., \& {Phillips}, T.~G.
  1992, \apj, 391, 121

\bibitem[{{Quillen} {et~al.}(2010){Quillen}, {Neumayer}, {Oosterloo}, \&
  {Espada}}]{2010PASA...27..396Q}
{Quillen}, A.~C., {Neumayer}, N., {Oosterloo}, T., \& {Espada}, D. 2010, \pasa,
  27, 396

\bibitem[{{Quillen} {et~al.}(2008){Quillen}, {Bland-Hawthorn}, {Green},
  {Smith}, {Prasad}, {Alonso-Herrero}, {Cleary}, {Brookes}, \&
  {Lawrence}}]{2008MNRAS.384.1469Q}
{Quillen}, A.~C., {et~al.} 2008, \mnras, 384, 1469

\bibitem[{{Rydbeck} {et~al.}(1993){Rydbeck}, {Wiklind}, {Cameron}, {Wild},
  {Eckart}, {Genzel}, \& {Rothermel}}]{1993AA...270L..13R}
{Rydbeck}, G., {Wiklind}, T., {Cameron}, M., {Wild}, W., {Eckart}, A.,
  {Genzel}, R., \& {Rothermel}, H. 1993, \aap, 270, L13

\bibitem[{{Sage} \& {Galletta}(1993)}]{1993ApJ...419..544S}
{Sage}, L.~J., \& {Galletta}, G. 1993, \apj, 419, 544

\bibitem[{{Sault} {et~al.}(1995){Sault}, {Teuben}, \&
  {Wright}}]{1995ASPC...77..433S}
{Sault}, R.~J., {Teuben}, P.~J., \& {Wright}, M.~C.~H. 1995, in ASP Conf. Ser.
  77: Astronomical Data Analysis Software and Systems IV, ed. R.~A. {Shaw},
  H.~E. {Payne}, \& J.~J.~E. {Hayes}, 433

\bibitem[{{Seigar} \& {James}(1998)}]{1998MNRAS.299..685S}
{Seigar}, M.~S., \& {James}, P.~A. 1998, \mnras, 299, 685

\bibitem[{{Sellwood}(2011)}]{2011MNRAS.410.1637S}
{Sellwood}, J.~A. 2011, \mnras, 410, 1637

\bibitem[{{Struve} {et~al.}(2010){Struve}, {Oosterloo}, {Morganti}, \&
  {Saripalli}}]{2010AA...515A..67S}
{Struve}, C., {Oosterloo}, T.~A., {Morganti}, R., \& {Saripalli}, L. 2010,
  \aap, 515, A67

\bibitem[{{van den Bergh}(1976)}]{1976ApJ...208..673V}
{van den Bergh}, S. 1976, \apj, 208, 673

\bibitem[{{van Gorkom} {et~al.}(1990){van Gorkom}, {van der Hulst}, {Haschick},
  \& {Tubbs}}]{1990AJ.....99.1781V}
{van Gorkom}, J.~H., {van der Hulst}, J.~M., {Haschick}, A.~D., \& {Tubbs},
  A.~D. 1990, \aj, 99, 1781
\bibitem[Young(2002)]{2002AJ....124..788Y} Young, L.~M.\ 2002, \aj, 124, 
788 
\bibitem[Young et al.(2011)]{2011MNRAS.414..940Y} Young, L.~M., Bureau, M., 
Davis, T.~A., et al.\ 2011, \mnras, 414, 940 

\end{thebibliography}
\end{document}